\documentclass{article}
\usepackage{spconf,amsmath,graphicx}
\usepackage{color}
\usepackage{amsfonts}
\usepackage{bm}
\usepackage{blindtext}
\usepackage{svg}
\usepackage{multirow}
\usepackage{placeins}
\usepackage{hyperref}

\def\blfootnote{\xdef\@thefnmark{}\@footnotetext}

\graphicspath{{figures/}}
\def\rts{\textit{RT}$_{60}$}

\title{SS-BRPE: Self-Supervised Blind Room Parameter Estimation Using Attention Mechanisms}

\name{$
\begin{array}{ccc}
	\mbox{Chunxi Wang$^{1}$, Maoshen Jia$^{1}$, Meiran Li$^{1}$, Changchun Bao$^{1}$, Wenyu Jin$^{2}$}.
\end{array}
$}

\address{
$^{1}$  School of Information Science and Technology, Beijing University of Technology, Beijing, China\\
$^{2}$ AcousticDSP Consulting LLC, St Paul, MN, United States
} 

\begin{document}
%\ninept
\ninept
\def\baselinestretch{.903}\let\normalsize\small\normalsize
\maketitle
\begin{abstract}
In recent years, dynamic parameterization of acoustic environments has garnered attention in audio processing. This focus includes room volume and reverberation time (\rts), which define local acoustics independent of sound source and receiver orientation. Previous studies show that purely attention-based models can achieve advanced results in room parameter estimation. However, their success relies on supervised pretrainings that require a large amount of labeled true values for room parameters and complex training pipelines. In light of this, we propose a novel Self-Supervised Blind Room Parameter Estimation (SS-BRPE) system. This system combines a purely attention-based model with self-supervised learning to estimate room acoustic parameters, from single-channel noisy speech signals. By utilizing unlabeled audio data for pretraining, the proposed system significantly reduces dependencies on costly labeled datasets. Our model also incorporates dynamic feature augmentation during fine-tuning to enhance adaptability and generalizability. Experimental results demonstrate that the SS-BRPE system not only achieves more superior performance in estimating room parameters than state-of-the-art (SOTA) methods but also effectively maintains high accuracy under conditions with limited labeled data. Code available at \url{https://github.com/bjut-chunxiwang/SS-BRPE}.
{\let\thefootnote\relax\footnote{{\noindent This work was supported by the National Natural Science Foundation of China under Grant No. 62471012 and Beijing Natural Science Foundation (No.L233032, L223033).}}}
\end{abstract}

\section{Introduction}
\vspace{-1mm}
Dynamic characterization of acoustic environments has garnered significant attention within the field of audio processing in recent years. Understanding parameters that define local rooms or acoustic spaces could be beneficial for a wide range of audio enhancement applications, including speech dereverberation, word recognition improvements for ASR and voice communication \cite{wu2017end,zhang2018deep}. Additionally, spatial sound reproduction systems \cite{cecchi2017room,jin2015theory} can utilize this data for tasks such as acoustic room equalization, thereby optimizing overall audio performance. In augmented reality (AR) applications, analyzing room acoustic parameters is also instrumental in generating perceptually acceptable sound, thereby ensuring a high-quality immersive experience \cite{Saini23}.

Given that environmental acoustic parameters and geometric information are closely linked to room impulse responses (RIRs), measuring RIRs can provide insights into factors such as reverberation time (\rts) and the direct-to-reverberant ratio (DRR). RIRs can also reveal other key parts of the so-called ``reverberation fingerprint", which includes location-independent parameters such as the geometric room volume. However, obtaining in-situ RIRs of a local acoustic environment is often challenging in practice due to the difficulties associated with implementing intrusive measurements \cite{jin2016adaptive}.

With advancements of deep learning techniques, using convolutional neural networks (CNNs) combined with time-frequency representations to address blind room acoustic parameter estimation from speech recordings as a supervised regression problem has become increasingly prevalent. CNN-based models demonstrated promising results in tasks involving blind estimation of \rts\ \cite{gotz2022blind,bryan2020impulse,gamper2018blind} and room volume \cite{genovese2019blind,ick2023blind}, as well as joint systems \cite{srivastava2021blind} that simultaneously estimate a set of room acoustic parameters in addition to \rts\  and volume, including total surface area, mean surface absorption, clarity, etc. 
By integrating with recurrent layers, CNNs can be extended into convolutional recurrent neural networks (CRNNs) that leverage the temporal dependencies in data, thereby handling variable-length input sequences more effectively \cite{deng2020online}. Additionally, to better capture distant global context information, hybrid models that combine CNNs with self-attention mechanisms have demonstrated cutting-edge results in this task \cite{saini2023blind,wang2024berp}. Wang et al. \cite{wang2024exploring} took one step further and devised the first convolution-free, purely attention-based model for blind room parameter estimation. This model achieves state-of-the-art (SOTA) performance and more advantageous robustness when handling practical blind estimation problems, demonstrating the feasibility of eliminating the reliance on CNNs.

All above-mentioned studies directly estimate room acoustic parameters from microphone recordings in a supervised learning manner following data-driven methods, which implies that the diversity and scale of the training data are crucial for model performances. For example, the success of the purely attention-based model we previously proposed in \cite{wang2024attention} largely depends on the labeled ImageNet pretraining, as well as extensive room parameter labeled audio data. Purely attention-based models are generally more demanding in terms of training data than CNNs. The study in \cite{dosovitskiy2020image} indicates that vision transformers (ViTs) outperform CNNs only when the training data size exceeds 100 million samples. Meanwhile, RIR datasets with accurately labeled groundtruth room parameters are very limited (especially true for room volume), which poses significant challenges. Therefore, the core issue to address in this paper is how to effectively estimate room parameters without relying on the high-cost labeled ImageNet pretraining and limited RIR datasets.

Inspired by the work in \cite{gong2022ssast} that explores a self-supervised Audio Spectrogram Transformer, in this work we propose a purely attention-based Self-Supervised Blind Room Parameter Estimation (SS-BRPE) model that is capable of estimating geometric room volume and \rts\ from single-channel noisy speech signals. Our system employs Gammatone magnitude spectral coefficients along with low-frequency phase spectrogram as inputs. Using the attention mechanism in transformers, this approach facilitates the capture of long-range global context. In addition, by utilizing unlabeled audio data, the proposed model is pretrained with joint discriminative and generative masked spectrogram patch modeling to enhance the performance, while reducing its dependency on labeled room parameter data. Experimental results confirm that the proposed self-supervised framework significantly alleviates the reliance on extensive labeled data while its  blind room parameter estimation performance even surpassing the supervised ImageNet pretrained method.

\section{Model architecture}
In this section, we propose a novel SS-BRPE system. This system employs a self-supervised learning strategy, allowing a purely attention-based model to learn from unlabeled audio data, thereby eliminating the dependency on labeled room parameter data. Additionally, we propose a dynamic feature augmentation method. This method enables to directly process and enhance 2-D audio feature blocks in an online fashion during the fine-tuning stage, effectively improving its adaptability and generalizability to different data types.

\subsection{Self-Supervised Blind Room Parameter Estimation Model}

\subsubsection{Audio Spectrogram Transformer}

\begin{figure}[ht]
    \raggedright
    \includegraphics[height=5.00cm]{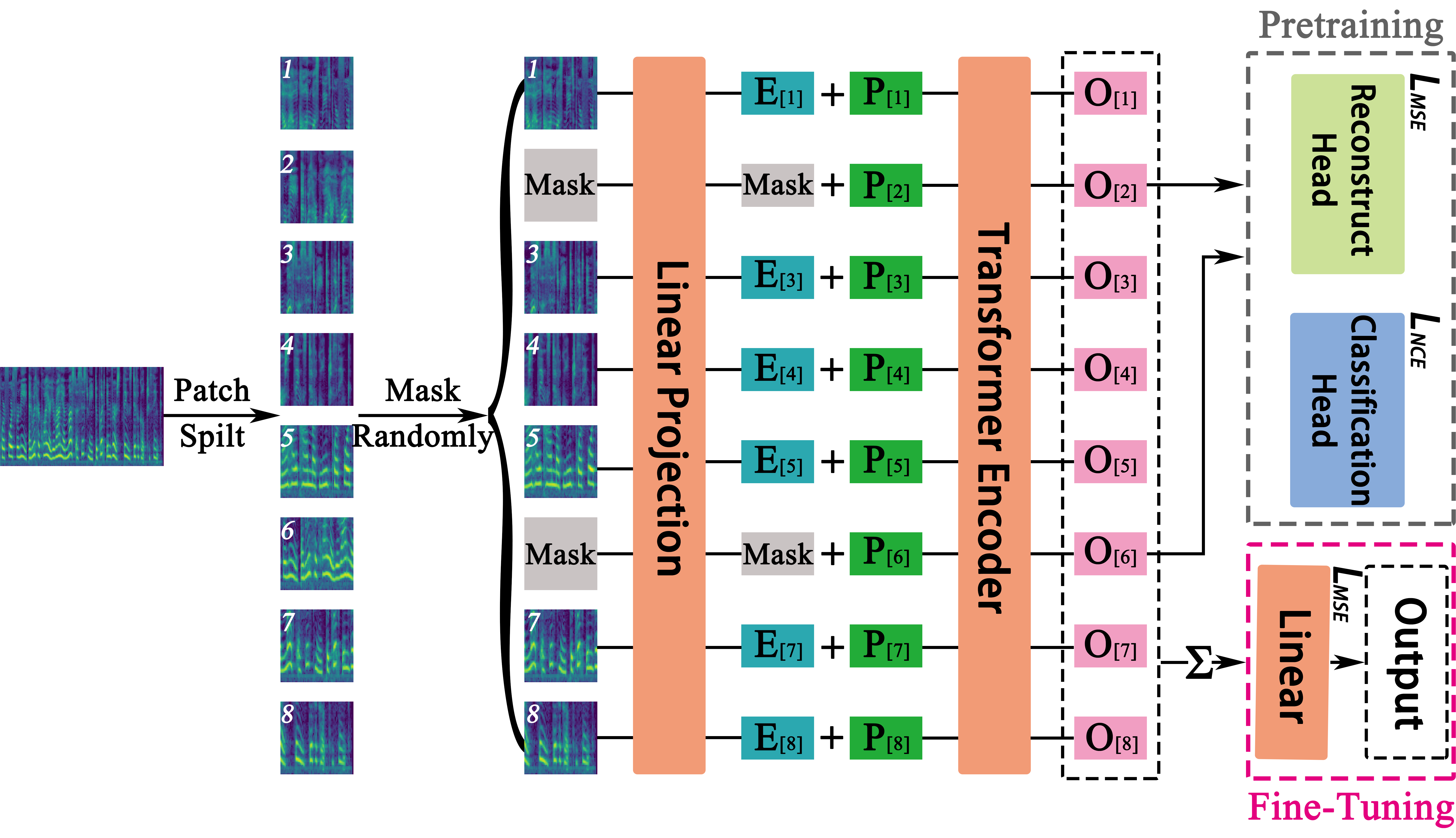}
    \vspace{-7mm}
    \caption{ The Self-Supervised Blind Room Parameter Estimation (SS-BRPE) architecture.}
    \label{fig:BERP}
    \vspace{-4mm}
\end{figure}

The proposed SS-BRPE system is depicted in Fig. \ref{fig:BERP}. The main body of the proposed system follows Audio Spectrogram Transformer (AST) \cite{gong2021ast} architecture. The audio is transformed into feature blocks and divided into $I$ patches, each size $16 \times 16$. The \(i\)-th patch $S_{[i]}$ is then flattened into a 1-D patch embedding of size 768 through a linear projection layer (referred as the patch embedding layer), resulting in embeddings denoted as $E_{[i]}$.

Since these patches are not arranged in chronological order and traditional Transformer architectures do not directly process the sequential order of input sequences, trainable positional embeddings $P_{[i]}$ with the same dimension of 768 are incorporated after each patch embedding. This allows the model to grasp the spatial structure of the audio spectrogram and understand the positional relationships among different patches. Furthermore, the combined embeddings ($E_{[i]}+P_{[i]}$) are processed by the Transformer encoder. The encoder's output denoted as $O_{[i]}$, is used as the spectrogram patch representation.

During fine-tuning and inference, we adjusted the input and output dimensions of the SS-BRPE system. Specifically, the input is a feature block containing room parameter information, while the output is the estimated room parameter label (volume or \rts). The output sequence of the patch embedding, $O_{[i]}$, is used as the feature representation of the 2-D audio feature block. Mean pooling is then applied to obtain the audio clip level representation, and a linear layer is used to estimate the room parameter labels.

Two necessary modifications were made to adapt the supervised AST architecture to the self-supervised learning framework. First, instead of using a \texttt{[CLS]} token for audio clip representation, we applied mean pooling over all patch representations. Second, we avoided overlapping splits of spectrogram patches during pretraining to prevent the model from leveraging overlapped edges as a shortcut for the task prediction, encouraging it to learn more meaningful representations. The patches were split with an overlap of 6 during fine-tuning and inference, same as \cite{wang2024attention}.

\subsubsection{Self-supervised Learning Framework}

Publicly available RIR datasets with labeled room parameter ground truth are extremely limited, posing significant challenges for blind room parameter estimation tasks. To address this issue, previously we attempted the following approaches: 1) a synthetic RIR dataset based on the image-source model; 2) labeled cross-modal transfer learning. Although the second method achieved notable results \cite{wang2024attention,wang2024exploring}, the supervised pretraining process based on ImageNet is highly complex, subject to constraints on limited similarity between vision and audio model architectures \cite{gong2022ssast}. Meanwhile, the demand for a substantial amount of labeled data still remains.

Compared to costly labeled data, unlabeled audio data is relatively easier to acquire. Therefore, in this work we attempt to construct a pretext task for room parameter estimation, utilizing unlabeled data to reduce the model's dependency on labeled data.

Specifically, the SS-BRPE system utilizes a self-supervised pretraining framework. During pretraining, input spectrograms are first divided into non-overlapping patches \( S_{[i]} \), and a portion of these patches is randomly masked. Embeddings of these masked patches are used as training targets, focusing on both discriminative and generative tasks. This method reinforces the model to learn the underlying structure of the audio data.

\subsubsection{Pretraining}

The discriminative objective concentrates on accurate identification of masked patches, using a classification head to output vectors. These are compared against embeddings of all other patches in the batch to calculate the InfoNCE loss \cite{oord2018representation}. On the other hand, the generative objective focuses on reconstructing the original content of the masked patches. Predictions are generated by a reconstruction head and evaluated using the mean squared error (MSE) loss. The total loss $L$ is a weighted sum of the discriminative (\(L_d\)) and generative (\(L_g\)) losses: \(L = L_d + \lambda L_g\), in which \(\lambda\) determines the relative contribution of each loss component and is set to 10 in this work.

For the self-supervised pretraining of the SS-BRPE system, we integrated and processed audio samples from two datasets, AudioSet-2M \cite{gemmeke2017audio} and LibriSpeech \cite{panayotov2015librispeech}. AudioSet-2M includes approximately 2 million diverse 10-second audio clips, while LibriSpeech provides 960 hours of English audiobooks. All audio sequences were standardized into a uniform duration of 10 seconds, downsampled to 16kHz, and converted to mono to ensure consistent training. Notably, these datasets do not contain any associated room parameter labels and focus solely on the audio components.

\subsection{Fine-tuning with Feature Augmention}
In the blind room parameter estimation task, noisy speech signals are transformed into 2-D time-frequency representations through a feature extraction process. This allows the model to be trained effectively and to capture information about the acoustic space efficiently. In this work, Gammatone ERB filterbank is used to transform audio signals into 2-D feature blocks that serve as input for the neural network. This method incorporates the ``+\textit{Phase}" model \cite{ick2023blind}, which leverages phase-related features and is shown to outperform methods that solely rely on amplitude-based spectral features. 

% The dimensions of the entire feature block are $30 \times 1997$, where 30 represents the feature dimension ${F}$, and 1997 represents the time dimension ${T}$.

Further, we explored how to enhance the generalizability of room parameter estimation models with a limited RIR dataset. In \cite{wang2024attention}, SpecAugment data augmentation method \cite{park2019specaugment} is utilized. Although this method expands the dataset, it also faces several challenges: 1) offline processing: data augmentation is conducted as an offline step prior to training, which increases the complexity of preprocessing and does not allow for updates to augmented data during training; 2) information loss: time/frequency random masking strips can lead to the loss of acoustic features contained in specific bands \cite{chang2024gap}, which is particularly critical for the task of blind room parameter estimation.

\begin{figure}
    \centering
    \includegraphics[height=4.00cm]{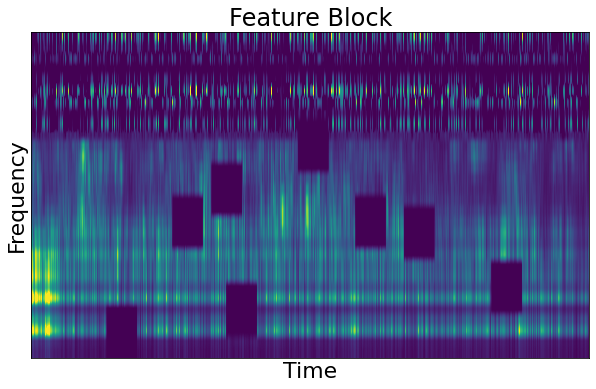}    
    \vspace{-3mm}
    \caption{Patch masking operation in online feature augmentation. Black patches represent random rectangular masks applied to 2-D audio feature blocks.}
    \label{fig:FA}
    \vspace{-4mm}
\end{figure}

To address these issues, we proposed a dynamic feature augmentation method in this work. Specifically, during the fine-tuning process, this method directly applies masking to the featurized 2-D audio feature blocks. Operating online, the method randomly selects 25\% of the samples in each batch for feature augmentation, ensuring diversity in the processed samples. For these selected samples, we implemented a rectangular patch masking operation based on 2-D features, as shown in Fig. \ref{fig:FA}. We established a random number and size of masking rectangles, randomly masking these blocks in each sample. This method not only enhances the diversity of features, enabling the model to better adapt to various types of input data in a dynamic manner during training, but also improves its generalizability without increasing data volume. 

\vspace{-1mm}
\section{Experiments}
\vspace{-1mm}
In this section, we evaluate the effectiveness of the proposed SS-BRPE system and compare it with the SOTA methods in the realm of single-channel blind room parameter estimation. First, the experimental design and setup of training sessions are introduced. Second, we present the estimation results of considered systems.

\vspace{-1mm}
\subsection{Experimental Design}
\begin{figure}
    \centering
    \includegraphics[height=7cm]{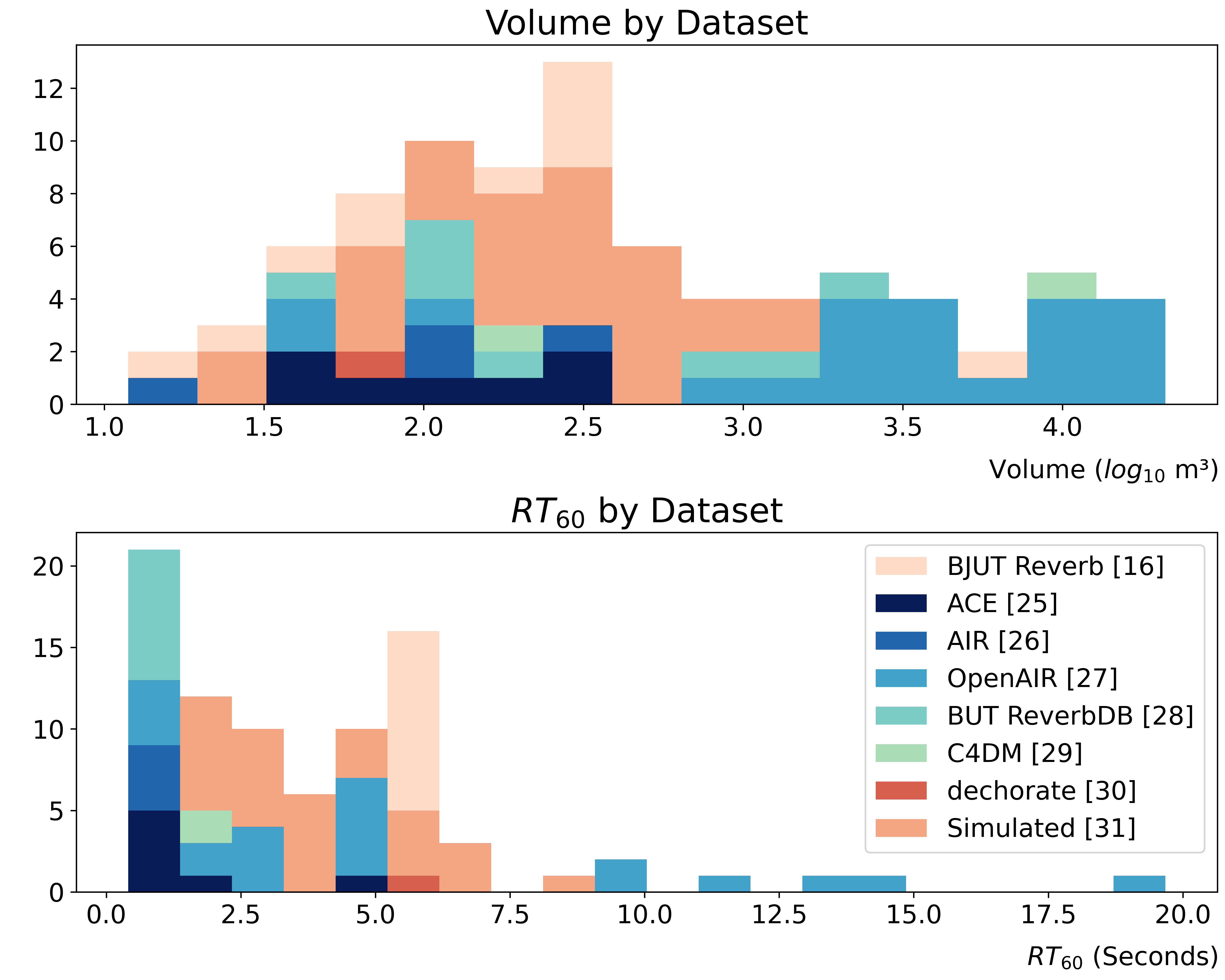}    
    \vspace{-3.5mm}
    \caption{Histograms of room volume and \rts\ distributions across various datasets. The horizontal axis represents scales of room parameters and the vertical axis represents numbers of rooms.}
    \label{fig:HIST}
    \vspace{-5mm}
\end{figure}

\vspace{-1mm}
We created an extensive audio sample library that encompassed a wide range of acoustic parameters using publicly available real-world RIR datasets \cite{wang2024attention, eaton2016estimation,jeub2009binaural, murphy2010openair,szoke2019building, stewart2010database,carlo2021dechorate} and a synthetic RIR dataset based on image source method \cite{scheibler2018pyroomacoustics}. Further, \rts\ values were measured using the Schroeder method \cite{kuttruff2016room}. All RIRs were uniformly downsampled to 16kHz. Room volume and \rts\ distributions across different datasets are illustrated in Fig. \ref{fig:HIST}. It is worth noting that for the test set, we selected only RIRs recorded in real-world environments to assess the model’s estimation performance on unseen non-simulated rooms.

To evaluate the performance of our SS-BRPE system, we compared it with the ``+\textit{Phase}" CNN-based model \cite{ick2023blind}, CRNN-based model \cite{wang2024exploring}, and a purely attention-based model with ImageNet pretraining \cite{wang2024attention}. We also added the feature augmentation scheme in
the fine-tuning process of the SS-BRPE system to verify its effectiveness. To accurately evaluate room parameters and address variability issues within smaller volume ranges, we applied base-10 logarithmic transformations to both volume and \rts, ensuring a more balanced weight across all acoustic space sizes during evaluation. During the model training phase, MSE was used as the loss function, and optimization was carried out using the Adam optimizer from PyTorch. CNN-based and CRNN-based models were trained for 1000 epochs, while the purely attention-based models underwent training for 150 epochs. This decision was based on observing good convergence behaviors during these epochs. To mitigate potential overfitting, L2 regularization was applied. Additionally, an adaptive learning rate strategy was employed to ensure effective convergence throughout the training process. If the model failed to show improvement on the validation set for ten consecutive epochs, early stopping criteria were applied to halt the training.

Four metrics on a logarithmic scale were used to assess the disparity between estimated and actual room parameters: MSE, Mean Absolute Error (MAE), Pearson correlation coefficient ($\rho$), and MeanMult (\textit{MM}). These statistical measures provide a comprehensive evaluation of both model accuracy and reliability. Additionally, median and MAE values on the linear scale were also reported to provides a more transparent insight of model performance.

In addition to the regular comparison study, we tested the performance of various models under limited data conditions, aiming to explore whether the SS-BRPE system can reduce its dependence on labeled room parameter data. This was achieved by randomly selecting 50\% of the room types to construct the training set while maintaining the integrity of the validation and test sets. This adjustment resulted in a reduction of 50\% in the number of audio samples, as well as the diversity of room types within the training set.

\begin{table*}[ht]
\caption{Performance comparison of the proposed SS-BRPE system with other supervised models.}
\vspace{+1mm}
\centering
\setlength{\abovecaptionskip}{5pt} % Reduce space above the caption
\setlength{\belowcaptionskip}{-5pt} % Reduce space below the caption (if needed)
\textbf{Volume Estimation}\\[0.1em]
\begin{tabular}{c|c|cccc|cc}
\hline
\multirow{2}{*}{\textbf{Method}} & \multirow{2}{*}{\textbf{Supervison}}&\multicolumn{4}{c|}{\textbf{Logarithmic Scale}} & \multicolumn{2}{c}{\textbf{Linear Scale}} \\
\cline{3-8}
& &MSE & MAE & $\rho$ & \textit{MM} & {Median} (m\textsuperscript{3}) & {MAE} (m\textsuperscript{3})\\
\hline
CNN \cite{ick2023blind} & Supervised &0.3863 & 0.4837 & 0.6984 & 3.0532 & 465.22 & 2239.12 \\
CRNN \cite{wang2024exploring} & Supervised & 0.3572 & 0.4265 & 0.7262 & 2.6701 & 371.70 & 2020.23 \\
Purely attention-based model w/ ImageNet \cite{wang2024attention} & Supervised & 0.2157 & 0.3111 & 0.8529 & 2.047 & 277.17 & 1735.16 \\
SS-BRPE & Self-supervised & 0.2003 & 0.2887 & 0.8937 & 1.9599 & 234.47 & 1532.32 \\
SS-BRPE w/ Feature AUG & Self-supervised & \textbf{0.1652} & \textbf{0.2721} & \textbf{0.8965} & \textbf{1.8773} & \textbf{223.69} & \textbf{1470.56} \\
\hline
\end{tabular}
%\vspace{1em}
\textbf{\rts\ Estimation}\\[0.1em]
\begin{tabular}{c|c|cccc|cc}
\hline
\multirow{2}{*}{\textbf{Method}} & \multirow{2}{*}{\textbf{Supervison}}&\multicolumn{4}{c|}{\textbf{Logarithmic Scale}} & \multicolumn{2}{c}{\textbf{Linear Scale}} \\
\cline{3-8}
& &MSE & MAE & $\rho$ & \textit{MM} & {Median} (s) & {MAE} (s)\\
\hline
CNN \cite{ick2023blind} & Supervised &0.1473 & 0.2966 & 0.8817 & 1.9952 & 0.25 & 1.9cm0 \\
CRNN \cite{wang2024exploring} & Supervised & 0.1068 & 0.2162 & 0.9235 & 1.9cm478 & 0.14 & 0.73 \\
Purely attention-based model w/ ImageNet \cite{wang2024attention} & Supervised & 0.0607 & 0.1824 & 0.9660 & 1.4556 & 0.12 & 0.52 \\
SS-BRPE & Self-supervised & 0.0479 & 0.1470 & 0.9633 & 1.4029 & 0.09 & 0.49 \\
SS-BRPE w/ Feature AUG & Self-supervised & \textbf{0.0370} & \textbf{0.1312} & \textbf{0.9720} & \textbf{1.3529} & \textbf{0.08} & \textbf{0.39} \\
\hline
\end{tabular}

\label{tab:per}
\end{table*}

\begin{table*}[ht]
\vspace{-5mm}
\caption{Performance comparison of various models under limited labeled RIR data conditions. }
\vspace{+2mm}
\centering
\setlength{\abovecaptionskip}{5pt} % Reduce space above the caption
\setlength{\belowcaptionskip}{-5pt} % Reduce space below the caption (if needed)
\begin{tabular}{c|c|c|c|c|c}
\hline
\textbf{Method} & \textbf{Estimation Type} & MSE & MAE & $\rho$ & \textit{MM} \\ \hline
\multirow{2}{*}{CNN \cite{ick2023blind}} & Volume & $0.4553 \pm 0.0070$ & $0.5395 \pm 0.0124$ & $0.6317 \pm 0.0083$ & $3.4623 \pm 0.0949$ \\ 
& \rts & $0.1959 \pm 0.0034$ & $0.3386 \pm 0.0042$ & $0.8379 \pm 0.0036$ & $2.1649 \pm 0.0216$ \\ \hline
\multirow{2}{*}{CRNN \cite{wang2024exploring}} & Volume & $0.4303 \pm 0.0099$ & $0.4877 \pm 0.0046$ & $0.6544 \pm 0.0090$ & $3.0747 \pm 0.0324$ \\ 
& \rts & $0.1653 \pm 0.0024$ & $0.3108 \pm 0.0055$ & $0.8657 \pm 0.0042$ & $2.0521 \pm 0.0265$ \\ \hline
{Purely attention-based model} & Volume & $0.2962 \pm 0.0056$ & $0.3968 \pm 0.0084$ & $0.7822 \pm 0.0097$ & $2.4411 \pm 0.1418$ \\ 
{w/ ImageNet \cite{wang2024attention}}& \rts & $0.0823 \pm 0.0020$ & $0.2119 \pm 0.0037$ & $0.9370 \pm 0.0028$ & $1.9cm290 \pm 0.0140$ \\ \hline
\multirow{2}{*}{SS-BRPE} & Volume & $0.2691 \pm 0.0035$ & $0.3527 \pm 0.0067$ & $0.8144 \pm 0.0044$ & $2.2564 \pm 0.0343$ \\ 
& \rts & $0.0671 \pm 0.0029$ & $0.1840 \pm 0.0051$ & $0.9477 \pm 0.0031$ & $1.5532 \pm 0.0386$ \\ \hline
\multirow{2}{*}{SS-BRPE w/ Feature AUG} & Volume & $\textbf{0.2247} \pm \textbf{0.0062}$ & $\textbf{0.3280} \pm \textbf{0.0079}$ & $\textbf{0.8413} \pm \textbf{0.0038}$ & $\textbf{2.1283} \pm \textbf{0.0388}$ \\ 
& \rts & $\textbf{0.0453} \pm \textbf{0.0021}$ & $\textbf{0.1492} \pm \textbf{0.0048}$ & $\textbf{0.9664} \pm \textbf{0.0005}$ & $\textbf{1.3970} \pm \textbf{0.0240}$ \\ \hline
\end{tabular}

\vspace{-4mm}
\label{table:limit}
\end{table*}

\subsection{Experimental Results}

\subsubsection{Estimation of Room Volume \& \rts}
We compared the performance of the SS-BRPE system with other models in volume and \rts\ estimation tasks separately. The goal of this experiment is to observe if we can match previous supervised training room parameter estimation models using a self-supervised learning approach, without extensive pretraining on labeled data. Experimental results are presented in Table \ref{tab:per}. 

It can be seen that the purely attention-based method significantly surpasses CNN-based and CRNN-based models. This demonstrates that fully attention-based neural network models are more efficient in terms of accurately learning and predicting room acoustic characteristics, even with the low-layer network configuration and a relatively small number of training epochs. This also corroborates our previous research findings \cite{wang2024attention}. In fact, results in \cite{wang2024attention} suggest that the large amount of labeled data in ImageNet pretraining facilitates a superior performance. Within the same framework of attention-based networks, we compared the SS-BRPE system, and a supervised model with ImageNet pretraining. Experimental results show that the proposed SS-BRPE demonstrates more superior performance in terms of prediction accuracy, relationship with ground truth values, and predictive capability. Furthermore, the deployment of the dynamic feature augmentation method elevated the performance of the SS-BRPE system to a new level, significantly improving the accuracy of room parameter estimation. As a more illustrative example, the test set includes room volumes ranging from 12 to 21,000 $m^3$, with \rts\ values between 0.41 and 19.68 seconds. The ``SS-BRPE w/ Feature AUG" system exhibited a median and MAE of only 223.69 $m^3$ and 1470.56 $m^3$, respectively, on the linear scale. The median and MAE for \rts\ values were 0.08 seconds and 0.39 seconds, respectively. 

These results indicate that the SS-BRPE system, through its self-supervised learning approach, effectively captures the intrinsic characteristics of the room parameter regression problem, demonstrating more superior model performance over supervised learning methods. More importantly, this learned model successfully generalizes to unseen real-world rooms.

\subsubsection{Estimation of Room Parameters with Limited Data}

To confirm whether the self-supervised learning method can maintain its performance with limited labeled RIR data,  estimation results of various models in estimating volume and \rts\ are shown in Table \ref{table:limit}. To ensure the reliability of results, we conducted the experiment five times and calculated the 95\% confidence interval for the outcomes.

 Under conditions of insufficient labeled RIR data, the performance of all models inevitably declines. The ``Purely attention-based model w/ ImageNet" method holds up reasonably well loss due to insufficient labeled RIR data, but it comes at the cost of requiring a substantial amount of labeled ImageNet data and a complex pipeline in supervised pretraining. These limitations restrict the practicality of purely attention-based models in room parameter estimation tasks. In contrast, the proposed ``SS-BRPE" system effectively mitigates performance degradation due to limited labeled RIR data and maintains high performance in blind room parameter estimation. Furthermore, by incorporating feature augmentation, the estimation accuracy is further improved, especially this improvement is even more pronounced when limited RIR datasets are available.
 
\section{Conclusion and Future work}
This paper proposes a SS-BRPE system enhanced by an attention mechanism and self-supervised learning. The system excels at estimating the geometric volume and \rts\ parameters of a room using unlabeled audio data for pretraining. This approach significantly improves the accuracy of blind room parameter estimation without relying on high-cost labeled data and ImageNet pretraining. Experimental results demonstrate that the SS-BRPE system performs excellently in single-channel blind room parameter estimation tasks, maintaining high performance even with limited data. Through dynamic feature augmentation, our model further enhances adaptability and generalization capabilities. Overall, this method provided an efficient and low-cost solution for blind room parameter estimation, showcasing its potential to accurately estimate indoor parameters in complex acoustic environments. In future research, we will continue to explore blind room parameter estimation algorithms based on more advanced models.

\bibliographystyle{IEEEbib}
\bibliography{refs}

\end{document}